\documentclass[preprint,12pt,sort&compress]{elsarticle}

\usepackage{amssymb}
\usepackage{srcltx}

\journal{Applied Surface Science}

\begin{document}

\begin{frontmatter}



\title{Hydrodynamic  approach to surface pattern formation by ion beams}


\author{Mario Castro}
%\email[]{}
\address{Grupo Interdisciplinar de Sistemas Complejos (GISC) and Grupo de Din\'amica No Lineal (DNL), Escuela T\'ecnica
Superior de Ingenier{\'\i}a (ICAI), \\ Universidad Pontificia Comillas, E-28015
Madrid, Spain}
\author{Rodolfo Cuerno}
%\email[]{}
\address{Departamento de Matem\'aticas and GISC, Universidad Carlos III de Madrid, Avenida de la
Universidad 30, E-28911 Legan\'es, Spain}

\begin{abstract}

On the proper timescale, amorphous solids can flow. Solid flow can be observed macroscopically in glaciers or lead pipes, but it can also be artificially enhanced by creating defects. Ion Beam Sputtering (IBS) is a technique in which ions with energies in the 0.1 to 10 keV range impact against a solid target inducing defect creation and dynamics, and eroding its surface leading to formation of ordered nanostructures. Despite its technological interest, a basic understanding of nanopattern formation processes occurring under IBS of amorphizable targets has not been clearly established, recent experiments on Si having largely questioned knowledge accumulated during the last two decades. A number of interfacial equations have been proposed in the past to describe these phenomena, typically by adding together different contributions coming from surface diffusion, ion sputtering or mass redistribution, etc.\ in a non-systematic way. Here, we exploit the general idea of solids flowing due to ion impacts in order to establish a general framework into which different mechanisms (such as viscous flow, stress, diffusion, or sputtering) can be incorporated, under generic physical conservation laws. As opposed to formulating phenomenological interfacial equations, this approach allows to assess systematically the relevance and interplay of different physical mechanisms influencing surface pattern formation by IBS.

\end{abstract}


\end{frontmatter}

% \linenumbers


%\maketitle

\section{Introduction}
\label{sec:intro}

\begin{figure}[!ht]
\begin{center}
\includegraphics[width=0.35\textwidth,clip=]{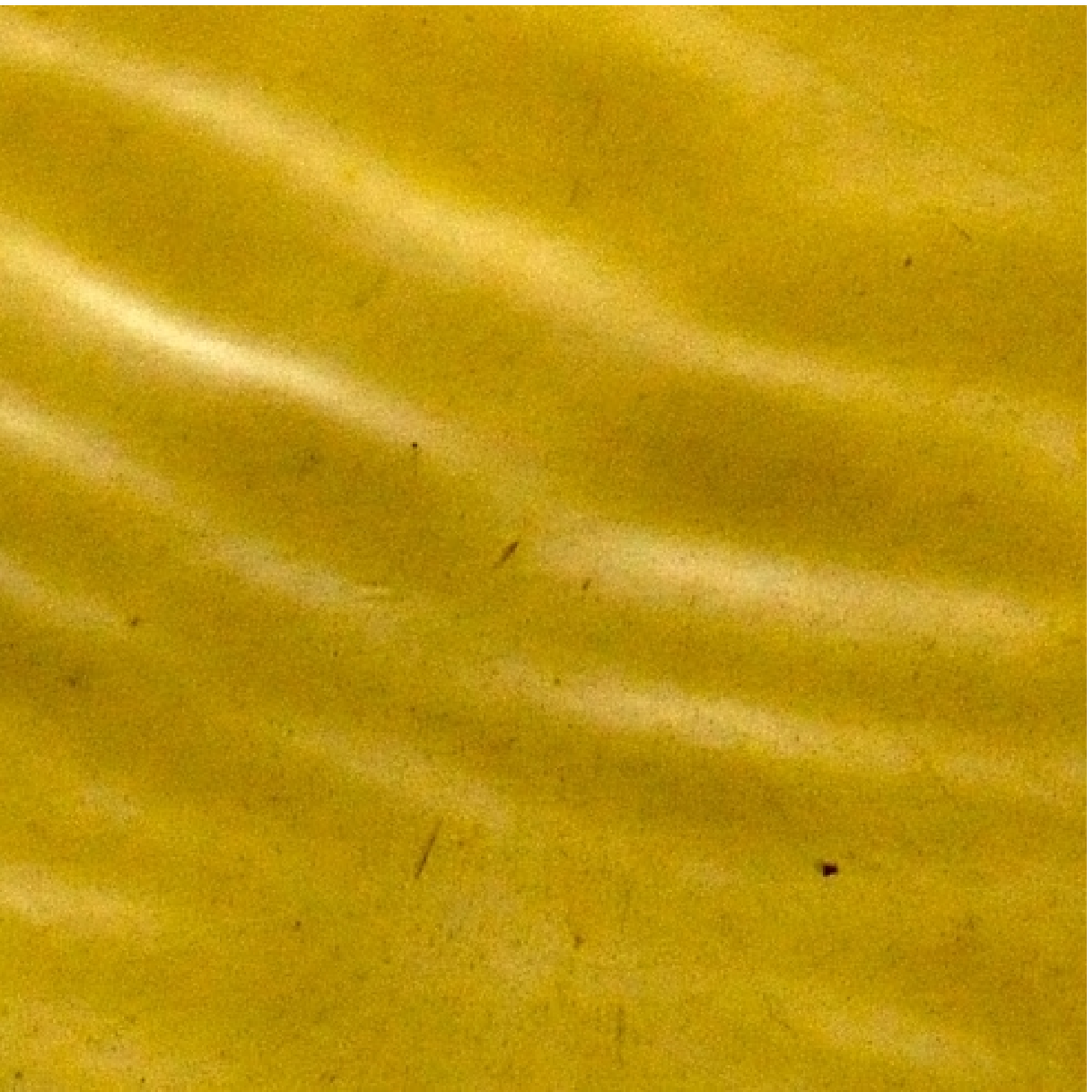}\\
\includegraphics[width=0.35\textwidth,clip=]{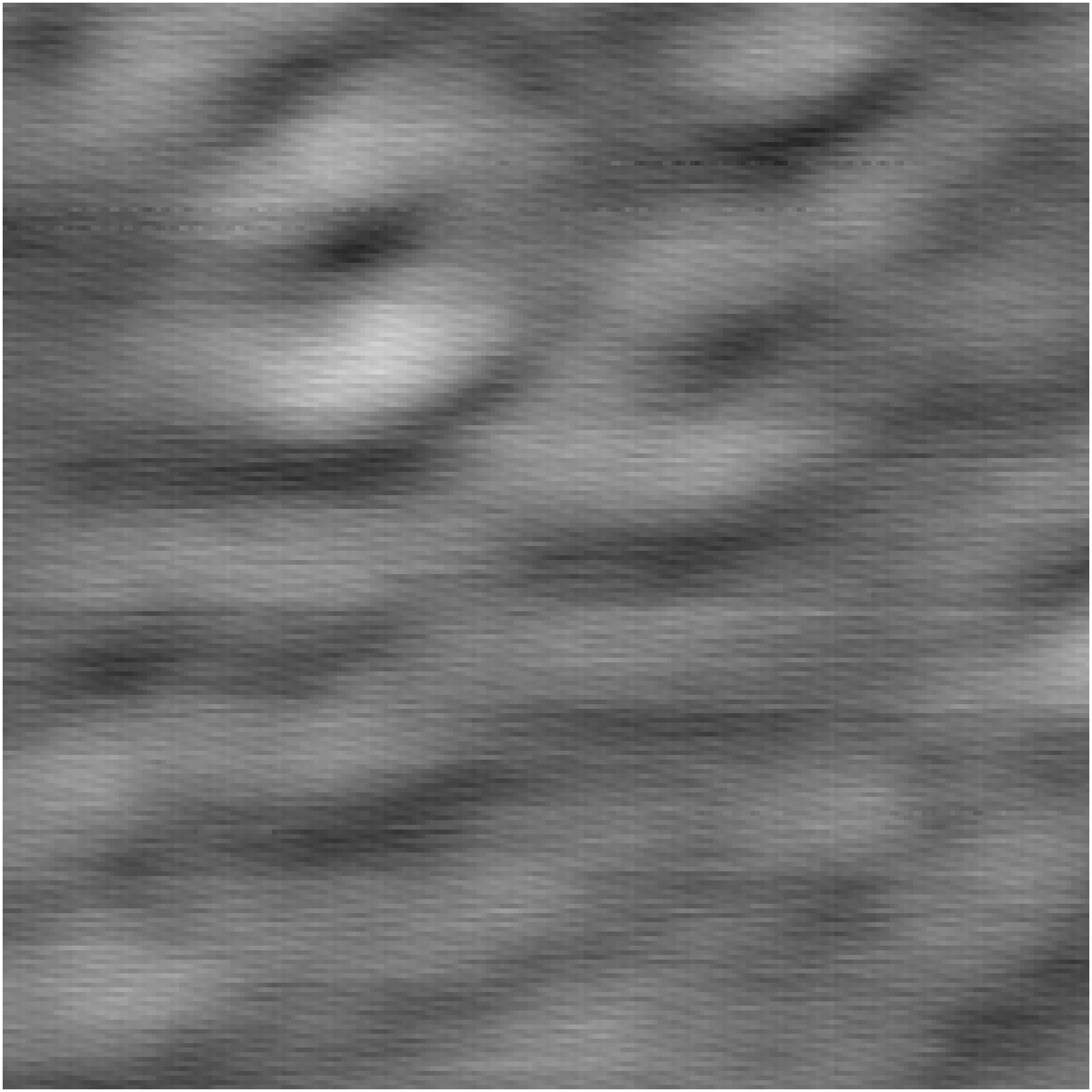}
\end{center}
\caption{\label{ripples} Water ripples (top) of characteristic wavelength $1$ cm vs Ion Beam Sputtering (IBS) ripples (bottom) on Silicon. Water ripples were obtained by shaking a photography bucket filled with water (coutesy of R. Vida, Universidad Pontificia Comillas).  Silicon ripples were obtained by bombarding Silicon with Ar$^+$. The width of the observation window is $1$ $\mu$m$^2$ (courtesy of L. V\'azquez, Instituto de Ciencia de Materiales de Madrid-CSIC).}
\end{figure}

Observations of nano-scale patterns on the surfaces of solid targets that undergo ion-beam sputtering (IBS) by 100 eV to 10 keV ions, date back at least to the early 1960's \cite{cunningham:1960,navez:1962}. They correspond either to amorphous materials like glass, or to targets that are amorphized by this type of irradiation, like semiconductors \cite{gnaser:1998}. The fascinating resemblance with macroscopic structures, like ripples on water or sandy dunes,\footnote{In Fig.\ \ref{ripples} we show the nice naked-eye similarity between water and Silicon ripples at very different orders of magnitude.} emphasized in \cite{navez:1962}, goes beyond a mere visual analogy \cite{munoz-garcia:2010,chan:2007_et_al}, underscoring the basic interest of this type of structures. Moreover, the high efficiency of IBS to produce surface nanopatterns (ripples and dots) over large areas on top of a wide variety of targets including metals and insulators %, and with a high degree of reproducibility
\cite{chan:2007_et_al,munoz-garcia:2009}, furnishes the technique with a large potential for applications.

%These are the type of targets that we will be considering in the present work.
Specifically, in IBS both the erosive process and the appearance of the patterns are due to the complex ion-target interactions: thus, ions create permanent defects in a surface layer whose thickness is of the order of the average penetration depth, trigger material ejection (sputtering) from its surface and induce material redistribution and flow \cite{volkert:1993,mayr:2003}. A conceptual framework explaining pattern formation in amorphous or amorphizable systems is available since the late 80's, through the work by Bradley and Harper (BH) \cite{bradley:1988} that was based on Sigmund's theory of linear collision cascades. Physically, a morphological instability \cite{cross:2009} originates in the higher erosion rate (yield) at surface minima as compared with surface maxima, in such a way that a pattern forms for any value of the angle of incidence and ion energy. This has become the theoretical paradigm supporting most experimental observations. Minor discrepancies with this view have been attributed to secondary effects, leading to refinements of BH's theory (see reviews in \cite{chan:2007_et_al,munoz-garcia:2009}).

However, many of the experiments done on elemental targets have been shown to be affected by contamination \cite{ozaydin:2005}, suggesting a correlation between pattern formation and a sizeable concentration of undesired species. In view of this, very recently clean experiments have been specifically tailored in order to avoid this feature
\cite{madi:2008,madi:2009,macko:2010,macko:2011}, taking Si targets as representative cases for the large class of substrates that become amorphous under irradiation. Remarkably, the outcome of these works invalidates the BH paradigm, since various morphological transitions are observed between unstructured and patterned surfaces as a function of both incidence angle $\theta$ and ion energy $E$,
confirming earlier observations at higher energies \cite{carter:1996} and contradicting one of the main predictions of BH theory. Thus, ironically, while for the case of multiple-component systems differential sputtering and species segregation may have been recently identified \cite{shenoy_et_al,leroy:2010,bradley:2010,zhou:2010} as the main physical mechanisms behind the morphological instability, the {\em a priori} simpler case of single-component systems still remains to be understood.

In addition, none of the current phenomenological models that elaborate on the BH paradigm~\cite{makeev:2002,facsko:2004,davidovitch:2007} are able to capture the complex morphological features elucidated by the recent experiments on Si. In these theories, the dynamics of the target surface is described through an effective evolution equation in which the various physical mechanisms are combined {\em ad-hoc}, and defect dynamics are neglected. The same limitations when confronted with experiments occur even for improved physical models in which the evolution of the surface height is explicitly coupled to that of the density of defects whose transport is confined to a {\em thin} surface layer, akin to pattern formation on e.g.\ aeolian sand dunes, see e.g.\ \cite{castro:2005,munoz:2006,munoz:2008,munoz:2009b}, and \cite{cuerno:2011} for a recent overview.

In order to overcome the limitations of previous continuum approaches to surface dynamics by IBS, we propose a generic framework based on general conservation laws (such as those governing mass or momentum), that allows to study systematically the role of the different mechanisms involved in the process. This approach, reminiscent of classical hydrodynamics, will allow us to identify the physics underlying the patterning of the eroded target surface. In summary, the main contribution of this work is to present a general consistent physical framework to study this system, that inherits the benefits of traditional fluid mechanics problems for which it has been proved highly successful in the past.

\section{Governing equations}
\label{ge}

In the energy range considered, an impinging ion interacts mostly with the nuclear structure of the target, generating vacancies and interstitials. Ions and defects can diffuse at different rates, annihilate by recombination or form clusters. As borne out from Molecular Dynamics (MD) simulations, the overall effect of these processes is threefold: amorphizing \cite{pelaz:2004}, stressing the target \cite{kalyanasaundaram:2008}, and displacing the mean atom position inside the material \cite{moseler:2005}. Moreover, as also seen in experiments \cite{gnaser:1998}, an amorphous layer forms rapidly \cite{moore:2004,kalyanasaundaram:2008}, its thickness and other mechanical properties like density, $\rho$, becoming stationary after a fluence of order $10^{14}$ ions cm$^{-2}$ (corresponding to a few seconds of irradiation for a typical ion flux of $10^{13}$ to $10^{15}$ ions cm$^{-2}$ s$^{-1}$). %, the system evolving quasi-statically.
Such a small ion flux drives dynamics slowly, inducing (as in macroscopic solid flow for e.g.\ a glacier) a time scale separation between atomistic relaxation rates ($\approx 1$ ps$^{-1}$ for collision cascades or adatom hopping), and the $\approx 10$ s times in which significant morphological changes occur. This fact legitimates a continuum description.

\subsection{Conservation of mass}

We are interested in the stages of the bombarding process in which the {\em amorphous layer} has already been formed. Thus, although its density value changes with respect to that for the pristine target, it does not evolve further after the initial transient just described. Mathematically, this fact can be expressed in the form
\begin{equation}
\nabla\cdot{\bf V}=0,
\label{contiunity}
\end{equation}
where ${\bf V}$ is the velocity field of the fluidized layer, whose components are given by
\begin{equation}
\mathbf{V} = u \mathbf{i}+ v \mathbf{j}+ w \mathbf{k}.
\label{V}
\end{equation}
In the language of fluid mechanics, the amorphous layer is said to be incompressible in the steady state.

\subsection{Conservation of momentum}

Conservation of momentum can be universally expressed as~\cite{oron:1997}
\begin{equation}
\rho D\mathbf{V}/Dt \equiv \rho (\partial_t{\bf V}+{\bf V}\cdot \nabla{\bf V}) =\nabla \cdot {\bf T},
\label{momentum}
\end{equation}
where ${\bf T}$ is the stress tensor. Indeed, the physics of the IBS problem enters the constitutive equation for this tensor.

Although a complete characterization of the stress tensor in the bombarded material is still lacking, there are strong evidences that the amorphous layer can be considered as a highly viscous fluid, hence ${\bf T }$ can be written as
\begin{equation}
T_{ij}=-P\delta_{ij}+\mu \left(\partial_{i}u_j+\partial_j u_i\right)+T^s_{ij},
\end{equation}
where $T^s_{ij}$ contains all the terms that do not come from hydrostatic pressure, $P$, or viscous flow, and $u_{1,2,3}\equiv u,v,w$ are the components of the velocity field (\ref{V}).

Moreover,  because the radiation induced viscosity, $\mu $, is around $15$ orders of magnitude larger than the viscosity of water \cite{mayr:2003}, we can safely drop the left hand side of Eq.\ (\ref{momentum}).  Thus, combining the conservation of mass with this assumption for the amorphous layer, we can write Eq.\ (\ref{momentum}) simply as
\begin{equation}
\mathbf{0} = \nabla\cdot {\bf T}^s+\mu  \nabla^2{\bf V}-\nabla P,
%\mathbf{0} = {\bf b}+\mu  \nabla^2{\bf V}-\nabla P,
\label{stokes}
\end{equation}
where we define ${\bf b}\equiv \nabla\cdot {\bf T}^s$ as a body force acting in the bulk of the fluid layer. Actually,
this force contains the relevant physics besides viscous flow. It is effective in the sense that detailed knowledge of the stress induced beneath the surface is not yet known experimentally. However, it is reasonable to expect that this force can be split into an amplitude, $f_E$,  and an angular contribution, $\Psi$, which is a function of the local angle of incidence. Mathematically,
\begin{equation}
{\bf b}={\bf f}_E \Psi(\theta -\gamma ), %= f_E \cos(\theta -\gamma )
\label{b_eq}
\end{equation}
where $\gamma$ is the local slope of the surface and $\theta $ is the incidence angle of the ion beam. Here, ${\bf f}_E$ contains the coarse-grained information about the effect of the residual stress created in the target, due to ion-induced mass redistribution, and has dimensions of a gradient of stress.

In three dimensions, and using standard spherical coordinates, we can write
\begin{eqnarray}
b_x&=&f_E\Psi(\theta -\gamma )\sin \theta \cos \phi, \\
b_y&=&f_E^\prime \Psi(\theta -\gamma )\sin \theta \sin \phi, \\
b_z&=&f_E\Psi(\theta -\gamma )\cos \theta,\label{bz}
\end{eqnarray}
where $\phi $ is the {\em azimuthal} angle (the angle in the plane of the target with respect to the plane of incidence of the ions). Possible anisotropies can enter the dynamics from the difference in the values of the amplitudes $f_E$ and $f_E^\prime$. We are assuming that the collision cascade that produces the damage and, ultimately, the strain/stress that drives the amorphous layer, is oriented in the direction of the incoming ion (see Fig. 2). The fact that $f_E$ is the same in the equations for $x$ and $z$ comes from the fact that there are, actually, just two important directions: parallel and perpendicular to the ion flux. When we project the parallel contribution in those directions ($x$ and $z$), the pre-factor remains the same.

\begin{figure}[!htp]
\begin{center}
\includegraphics[width=0.9\textwidth,clip=]{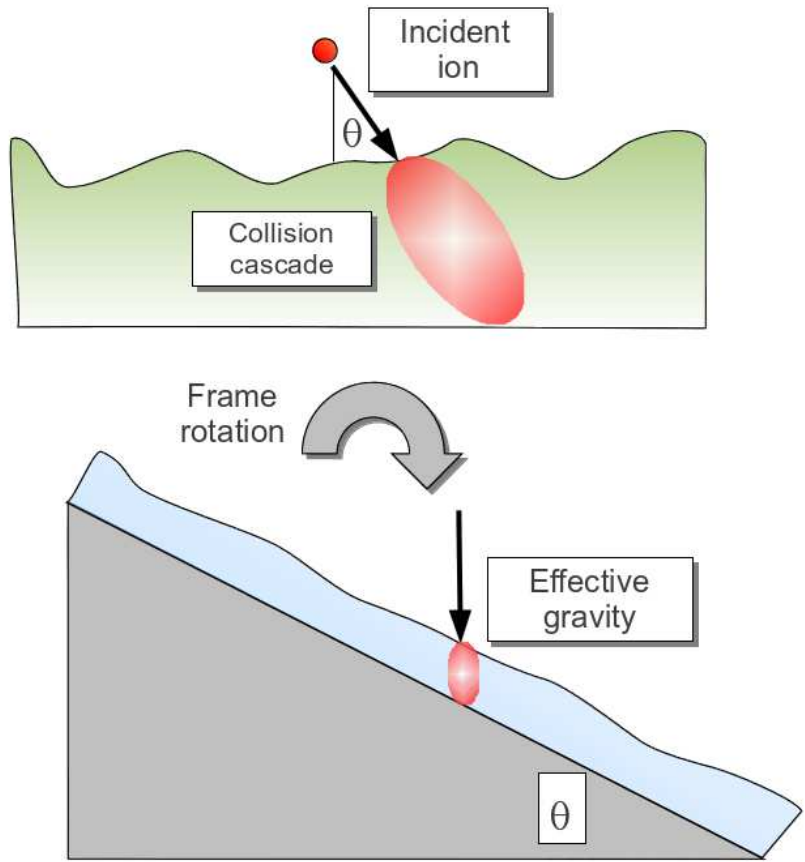}
\end{center}
\caption{\label{analogy}Analogy between IBS and fluid flow down a inclined plane. By rotating the frame of reference so that the ion impacts vertically, the IBS system can be seen as fluid flow down an inclined plane. The body force {\bf b} is equivalent to a non-homogeneous gravity field that depends on the local slope value at the free interface.}
\end{figure}

This body force can be understood as an effective gravity term. Thus, in Fig.\ \ref{analogy} we provide a cartoon to illustrate the analogy between ion induced solid flow and a genuine fluid flowing down an inclined plane. The two systems are mathematically equivalent once a rotation of angle $\theta$ is performed in the laboratory reference frame. The main difference between both systems comes from the corresponding gravity fields. While for the standard fluid problem gravity is a constant, for the erosion system the effective gravity changes from point to point on the surface, its value depending strongly on the local surface orientation with respect to the ion beam. Thus, unlike the fluid problem, in the erosive case one can obtain a morphological instability for appropriate angles of incidence.

\subsection{Boundary conditions}

Equations (\ref{contiunity})-(\ref{bz}) need to be supplemented with proper boundary conditions involving the erosive terms and the stress at the interfaces. In the most general case, we have to specify the evolution for two important boundaries of the system: the amorphous-vacuum interface (the free surface), $h^{(a)}$, and the amorphous-crystalline interface, $h^{(c)}$.

At the free surface, the balance of stress takes the form \cite{oron:1997}
\begin{eqnarray}
\mathbf{n}\cdot\mathbf{T}\cdot\mathbf{n}& =& - \sigma \kappa+T _n^{(ext)},\\
\mathbf{n}\cdot\mathbf{T}\cdot\mathbf{t}& =& \nabla\sigma  \cdot {\bf t}+T _t^{(ext)},\\
\mathbf{n}\cdot\mathbf{T}\cdot\mathbf{q}& =& \nabla\sigma  \cdot {\bf q}+T _q^{(ext)},
\end{eqnarray}
where $\mathbf{n}$ is the unit outward normal vector,  $\mathbf{t}$ and $\mathbf{q}$ are two mutually perpendicular vectors in the tangent plane to the surface, $\kappa$ is the mean surface curvature, and $\sigma$ is surface tension. Finally, $T_i^{(ext)}$ are the projections along the corresponding directions of an external stress that is applied over the surface.

Here we will consider that the surface tension is constant. However, for multicomponent materials where one species segregates at the surface, one can assume that there is a surface tension gradient that may affect the dynamics of the surface (the so-called Marangoni effect \cite{oron:1997,Kree}).

Finally, a kinematic condition~\cite{oron:1997} at each interface leads to evolution equations for both $h^{(a)}$ and $h^{(c)}$ as
\begin{equation}
\frac{Dh^{(c)}}{Dt}-w= j_{am},
\label{crystal}
\end{equation}
and
\begin{equation}
\frac{Dh^{(a)}}{Dt}-w= j_{er},
\label{erosionsurface}
\end{equation}
where, as defined above, $w$ is the vertical component of the velocity field.
The terms $j_{am}$ and $j_{er}$ stand for the rates of amorphization and erosion, respectively. Thus, Equations (\ref{crystal}) and (\ref{erosionsurface}) express mathematically that the difference between the vertical velocity field and the actual motion of the interface is equal to the rate at which material is removed from both phases.

For instance, the erosive (BH-type) mechanisms enter directly through $j_{er}$, so that their contribution to the full dispersion relation will be additive, see section \ref{la}. For simplicity, we assume that the rate of amorphization is equal to the rate of erosion, since in experiments both interfaces evolve at the same rates in the steady state. Moreover, for mathematical tractability, we assume that the crystalline-amorphous interface is flat. Under these assumptions, a final boundary condition is required, by which we specify that, at $h^{(c)}$, the fluid moves with the same velocity as the crystalline phase; this is often referred to as the no-slip boundary condition \cite{oron:1997}.

\section{Linear analysis}
\label{la}

The standard procedure to study the linear stability of a surface is developed in three steps:
\begin{enumerate}
\item Obtaining the flat (zeroth order) interface solution.
\item Making a periodic infinitesimal perturbation of the solution found in the previous step, with wavelength $\lambda =2\pi /q$ ($q$ being the wavenumber).
\item Calculating the amplification rate, $\omega _q$, for every wavenumber $q$ (also known as linear dispersion relation).
\end{enumerate}

For the sake of simplicity, we will detail these steps for the two dimensional case (namely, the interface is a line and not a proper surface). At the end of our derivation, we will quote the dispersion relation for the full three dimensional case.

\subsection{Flat interface solution}

We fix our reference frame on the free surface, so that the crystalline-amorphous interface is located at $z=-d$, where we denote by $d$ the average thickness of the amorphous layer and by $h(x)$ the fluctuations of the free surface. The flat interface condition is given simply by $h(x)=0$. Introducing it in the equations for the velocity field, we find that the flat solution is further characterized by
\begin{eqnarray}
P_0(z)&=&-f_Ez\cos\theta \Psi(\theta ),\label{P0}\\
u_0(z)&=&\frac{f_E(d^2-z^2)\sin\theta \Psi(\theta )}{2\mu },\\
w_0(z)&=&0.
\end{eqnarray}

\subsection{Linear perturbation}

To study the morphological stability of the flat interface, we look for solutions of the governing equations of the form
\begin{equation}
h(x)=\epsilon e^{\omega _qt-iqx},
\label{hepsilon}
\end{equation}
for an arbitrary wavenumber $q=2\pi /\lambda $. Similarly, we assume that the pressure and the velocity fields can be written as
\begin{eqnarray}
P(x,z)&=&P_0(z)+\epsilon P_1(z)e^{\omega _qt-iqx},\label{Pexpansion}\\
u(x,z)&=&u_0(z)+\epsilon u_1(z)e^{\omega _qt-iqx},\\
w(x,z)&=&w_0(z)+\epsilon w_1(z)e^{\omega _qt-iqx},\label{wexpansion}
\end{eqnarray}
and solve the governing equations perturbatively to linear order in the small parameter $\epsilon$.
The precise form of the ensuing corrections to the flat interface solution, $P_1(z)$, $u_1(z)$ and $w_1(z)$, is given by
Eqs.\ (\ref{P1})-(\ref{w1}) in \ref{app:a}, respectively.

\subsection{Dispersion relation}

Finally, the dispersion relation can be extracted from the kinematic condition. Specifically, from Eqs.\ (\ref{erosionsurface}) and (\ref{hepsilon}) we get
\begin{eqnarray}
 & & \partial _th =-u(x,0)\partial _xh+w(x,0)+j_{er} \nonumber \\
 & \rightarrow & \omega _q =iq  u_0(0)+ w_1(0)+\bar{\omega}_{q},
\end{eqnarray}
where $\bar{\omega}_{q}$ comes directly from the erosive contribution $j_{er}$; here is where, for instance, the classical theory of Bradley and Harper can enter~\cite{bradley:1988}.

Using Eqs.\ (\ref{P0})-(\ref{wexpansion}) and denoting real part with a single prime and imaginary part with a double prime,
we find:
\begin{equation}
\omega _q^\prime=-\frac{\bigg(q^2 \sigma +f_E\partial _\theta (\sin(\theta ) \Psi (\theta ))\bigg)(-2 d q+\sinh(2 d q)) }{2 q \mu  \bigg(1+2 d^2 q^2+\cosh(2 d q)\bigg)}+\bar{\omega}_{q}^\prime
\label{realwq}
\end{equation}
and
\begin{eqnarray}
&&\omega _q^{\prime\prime}=\frac{1}{2 q \mu  \bigg(1+2 d^2 q^2\cosh(2 d q)\bigg)} \bigg(
f_E \bigg(d^2 q^2 \bigg(3+2 d^2 q^2+\nonumber \\
&&  \cosh(2 d q)\bigg) \sin(\theta ) \Psi (\theta )-2 \cos(\theta ) \bigg(1+d^2 q^2-2 \cosh(d q)+\cosh(2 d q)-\nonumber \\
&&2 d q \sinh(d q)\bigg) \Psi '(\theta )\bigg)\bigg)+\bar{\omega}_{q}^{\prime\prime}.
\label{imagwq}
\end{eqnarray}

\subsubsection{Real part: Stability of the flat interface}

The real part of $\omega _q$ controls the stability of the flat interface: if $\omega^\prime _q$ is positive/negative, a periodic perturbation to the flat solution with wavenumber $q$ will grow/decay exponentially as given by Eq.\ (\ref{hepsilon}). 
Equation (\ref{realwq}) generalizes important results from fluid dynamics, and has already been reported in the context of the IBS problem for more specific choices of the function $\Psi$ \cite{arxiv,cuerno:2011}. For instance, for $f_E\rightarrow 0$, it reduces to Orchard's classic result \cite{orchard:1962} on the flow of a viscous layer of arbitrary thickness on top of an immobile substrate. In particular, $qd \gg 1$ then leads to Mullins' \cite{mullins:1959} celebrated rate of flattening when relaxation occurs through bulk viscous flow, $\omega _q=-\sigma q/2\mu$, already invoked for some IBS experiments \cite{chason:1994,frost:2009}.

Generally, equation (\ref{realwq}) is non-polynomial, implying non-local effects in the dynamics of the amorphous layer.
This non-locality may be important when other non-local effects are present in the dynamics (as redeposition, shadowing, or secondary ion scattering). Moreover, in our case the sign of $\omega _q^\prime$ may depend on the incidence angle, $\theta $, thus explaining morphological transitions from flat to patterned structures at different values of $\theta $ \cite{madi:2008,madi:2009,macko:2010,macko:2011}.

\subsubsection{Imaginary part: in-plane propagation of the pattern}
In those cases in which an unstable wavelength is selected (namely, $\omega _q^\prime>0$ for some $q$), the imaginary part of $\omega _q$ provides information about the velocity of lateral in-plane propagation of the pattern. Specifically, if the most unstable wavenumber is $q_c$ (namely, the one with the largest positive value of $\omega _q^\prime$), then the pattern propagates with velocity
\begin{equation}
V_{ripple}=\frac{d\omega _q^{\prime\prime}}{dq}\bigg|_{q=q_c}.
\end{equation}
Again, the sign of $V_{ripple}$ may depend on the incidence angle; thus, depending on the choice of $\Psi(\theta )$, the ripples can move in the same direction as or opposite to the ion beam.

\subsubsection{``{\em Shallow water}'' approximation ($d/\lambda \ll 1$)}

For large enough energies, and because the typical wavelengths of the patterns reported in the literature for energies around 1 keV are in the $\lambda \approx 20-100$ nm range, thus an order of magnitude larger than the amorphous layer thickness ($d\approx 1-5$ nm), we can assume that $qd\ll 1$ and Taylor expand $\omega^\prime _q$. In fluid mechanics this is known as the {\em shallow-water} approximation \cite{oron:1997}. Thus, we get
\begin{equation}
\omega^\prime _q=-\frac{f_E \partial_\theta (\Psi(\theta )\sin(\theta))d^3}{3\mu}q^2 -\frac{\sigma d^3}{3\mu }q^4+\overline{\omega}^\prime_q ,
\label{rewqshallow}
\end{equation}
where it is understood that a similar expansion has been done in the erosive contribution $\overline{\omega}^\prime_q$.
Note that, if we consider solid flow as the only pattern forming mechanism, we can simply drop this term. Since we are interested in probing the morphological consequences of viscous flow, we shall do this henceforth unless otherwise stated.

Equation (\ref{rewqshallow}) predicts that, if the function $\partial _\theta (\Psi(\theta )\sin\theta )$ is positive, the interface is stable (flat or rough, depending on the intensity of the thermal and beam flux fluctuations); on the contrary, if it is negative, a pattern appears with an angle dependent wavelength.

% Comment to referee

At this point, note that our hydrodynamical formulation is a coarse grained description of the microscopic phenomena. Thus, it has been recently shown that the ion impacts produce a response in the target referred to as a {\em prompt} regime~\cite{norris:2011}. In this reference, starting from atomistic molecular dynamics (MD) simulations, the overall effect of the ion impacts on the evolution of the surface height is summarized into the moments, $M^{(i)}$, of a so-called {\em crater function}, that eventually contribute to a continuum evolution equation for the surface height. It is worth noting that we can {\em map} the approach \cite{norris:2011} with our present formalism through
\begin{equation}
\Psi(\theta )=M_X^{(1)}/\tan(\theta ).
\label{}
\end{equation}
In this respect, we could incorporate MD information into our hydrodynamical description through our function $\Psi$. Let us
note incidentally that, as suggested by the simulations in \cite{norris:2011,kim:2011}, purely erosive contributions to the evolution of the surface height are numerically negligible, justifying our neglect of $\overline{\omega}^\prime_q$ above.

Patterns generated in the unstable parameter region in which Eq.\ (\ref{rewqshallow}) has a positive maximum for $q=q_c$, have a characteristic wavelength given by
\begin{equation}
\lambda_c = \frac{2\pi}{q_c} = 2\pi \left({\frac{2\sigma}{-f_E \partial_\theta (\Psi(\theta )\sin(\theta))}}\right)^{1/2}.
\label{critlambda}
\end{equation}
For instance, if one assumes that the angular correction is $\Psi(\theta)=\cos\theta $ (namely, that the body
force is proportional to the local flux at the surface), then Eq.\ (\ref{rewqshallow}) predicts a transition
from flat to patterned surfaces at $\theta =45^\mathrm{o}$ for all ion energies, as has been observed in experiments of
Ar$^+$ on Si at energies between $250$ eV and $1$ keV~\cite{madi:2009}. Similarly, for this choice of $\Psi$, the velocity of ripple is given by
\begin{equation}
V_{ripple}=\frac{f_Ed^2}{2\mu }\sin2\theta \left(1+\frac{3f_Ed^2\cos2\theta }{16\sigma} \right) .
\end{equation}

\begin{figure}[!ht]
\begin{center}
\includegraphics[width=0.49\textwidth,clip=]{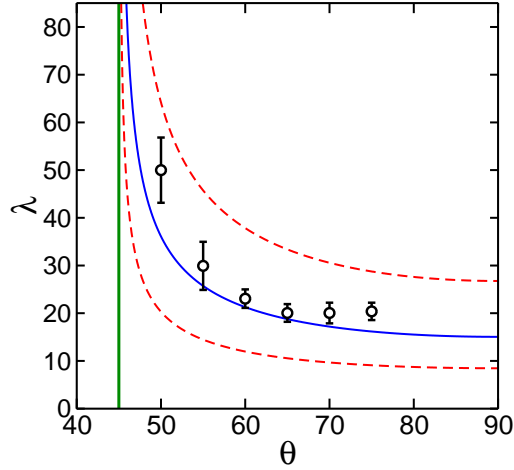}
\end{center}
\caption{\label{fig_nueva}
$\lambda_c(\theta)$ at the type II transition. Experimental data from \cite{madi:2008,madi:2009} are shown as black symbols, the green vertical line is the transition angle, and the blue solid line is the present theoretical prediction from Eq.\ (\ref{critlambda}); the red dashed lines provide our confidence interval, due to uncertainties in parameter values corresponding to experimental data. $\lambda $ is nm and $\theta $ in degrees.
}
\end{figure}

To be more quantitative, we need to determine the values of the main parameters of the theory. As mentioned above, the new parameter $f_E$ is related to the stress gradient created across the amorphous layer, induced by the collisional processes. There is a large experimental uncertainty in the measurement of stress, a property that depends moreover on the energy of the incident ions. Different authors report different values for ion-induced stress on Si, in the range $200$ MPa to $1.62$ GPa \cite{madi:2009,kalyanasaundaram:2008}. We are interested in the order of magnitude, hence we take the geometrical mean (which accounts for the {\em average} order of magnitude) of both extreme values, thus, approximately 569 MPa. This corresponds to $f_E=0.424$ kg nm$^{-2}$s$^{-2}$ for a layer of $d=3$ nm depth. In Fig.\ 3, this value corresponds to the blue solid line, while the limiting values of $200$ MPa and $1.62$ GPa are used to plot the red dashed lines, which we take as our confidence interval. More accurate measurements of stress would provide improved estimations. Additional parameters have not been measured experimentally yet. However, we can provide indirect estimates for their values when these are not explicitly available, as shown in Table 1. With these parameters at hand, Eq.\ (\ref{critlambda}) and experimental data are seen to agree quite closely in Fig.\ \ref{fig_nueva}.
\begin{table}[!h]
\begin{center}
\begin{tabular}{|c|c|c|}\hline
{\bf Parameter} &{\bf Value}&{\bf Reference}\\ \hline
Surface tension ($\sigma$)&$1.34$ J/m$^{2}$& \cite{vauth:2007} \\ \hline
Amorphous layer thickness ($250$ eV) &$3$ nm& \cite{madi:2008,madi:2009} \\\hline
Viscosity ($\mu$)& $6\times 10^8$ Pa s & \cite{norris:2011}$^{*}$\\ \hline
Stress range&$200$ MPa--$1.62$ GPa&\cite{madi:2009,kalyanasaundaram:2008}\\ \hline
\end{tabular}
\caption{Numerical values of the parameters used in the main text. $^{*}$Mayr et al.\ \cite{mayr:2003} report the value $\mu=10^{9}$ Pa s, which also leads to good agreement with the results presented in this work.}
\end{center}
\end{table}

The dependence of the pattern properties on other parameters such as energy, flux, etc., that enter the amplitude $f_E$, will be the subject of future work.

\subsection{Dispersion relation for the three dimensional case}

For completeness, we include here the dispersion relation for the three dimensional case. Now, the flat interface is a plane, and deviations to linear order take the form
\begin{equation}
h(x,y)=\epsilon e^{\omega_{{\bf q}}t-i{\bf q}\cdot{\bf x}},
\end{equation}
with ${\bf q}=q_x{\bf i}+q_y{\bf j}$ and ${\bf x}=x{\bf i}+y{\bf j}$. In the ``shallow-water'' limit, the dispersion relation is now given by
\begin{equation}
\omega _{{\bf q}}=-\frac{f_E^\prime \cos\theta \Psi(\theta )d^3}{3\mu }q_y^{2}-\frac{f_E\partial_\theta (\sin\theta \Psi(\theta ))d^3}{3\mu }q_x^2-\frac{\sigma d^3}{3\mu }(q_x^2+q_y^2)^2 + \overline{\omega}^\prime_q ,
\end{equation}
for a general angular dependence $\Psi$ of the body force and general (anisotropic) forms for the strain induced by the ion through the amplitudes $f_E$ and $f_E^\prime$.

\section{Connection with other continuum frameworks}
\subsection{Two-field ``hydrodynamic'' theory}

Despite the fact that we have not considered the dynamics of the amorphous-crystalline interface, here we want to note that, in the {\em shallow-water} limit, one can reduce the present full hydrodynamical formulation, making contact with previous theories of erosion, specifically with the two-field model approach pioneered in~\cite{aste} and further developed in \cite{castro:2005,munoz:2006,munoz:2008,munoz:2009b}, see an overview in \cite{cuerno:2011}.

Within this approach (see Ref.\ \cite{munoz:2008} for details), the dynamics of the density $R$ of moving species and that of the free surface height, $h$, are explicitly described, within the implicit assumption that the thickness of the surface layer where material transport takes place is negligible. The evolution of these two fields can be expressed through proper rate equations, borrowed from the field of pattern formation on aeolian sand dunes. Specifically,
\begin{eqnarray}
\partial_t R&=&(1-\phi )\Gamma_{ex}-\Gamma_{ad}+D\nabla^2R-{\bf V}\cdot\nabla R,\label{twofieldR}\\
\partial_t h&=&-\Gamma_{ex}+\Gamma_{ad},\label{twofieldh}
\end{eqnarray}
where $(1-\phi )$ measures the fraction of atoms that, having been dislodged from their equilibrium positions, are actually sputtered. Here, $\Gamma_{ex}$ and $\Gamma_{ad}$ are, respectively, rates of atom excavation from and addition to the (crystalline) immobile bulk.

In the hydrodynamical framework proposed in the present work, the kinematic conditions (\ref{crystal}) and (\ref{erosionsurface}) can be combined and reinterpreted within the two-field model approach. Thus, if we define
\begin{equation}
R= h^{(a)}-h^{(c)},
\end{equation}
then
\begin{equation}
\frac{DR}{Dt}=\partial_tR+{\bf V}\cdot\nabla R=j_{er}-j_{am}.
\end{equation}
Finally, we just need to define
\begin{eqnarray}
j_{er}&=&-\phi\Gamma_{ex}+D\nabla^2R,\\
j_{am}&=&-\Gamma_{ex}+\Gamma_{ad}.
\end{eqnarray}
to recover Eq.\ (\ref{twofieldR}) exactly.

\subsection{Lubrication theory}

The hydrodynamical nature of the model considered in Sections \ref{ge}, \ref{la} suggests a direct comparison with the so-called lubrication theory of fluids. This theory is valid in the shallow-water approximation which is, in fact, the one relevant to IBS experiments.

Following Oron {\em et al.} \cite{oron:1997}, we define a small parameter $\varepsilon=qd$. Additionally, we assume that not only the slopes are small but also the time-scales related to the motion of the fluid are slow (with respect to the motion of the very same surface). Mathematically, the following change of variables is defined:
\begin{eqnarray}
X=\varepsilon x/d, Z=z/d,U=u/U_0,W=w/\varepsilon U_0\textrm{ and }T=\varepsilon U_0t/d.
\end{eqnarray}
In order to capture the essential ingredients that affect the motion of the fluid, we rescale the stress tensor and the corresponding values at the boundary. Thus,
\begin{equation}
\tau_{xx}=\frac{d}{\varepsilon U_0}T_{xx}, \tau_{xz}=\frac{d }{U_0}T_{xz}, \tau_{zz}=\frac{d }{\varepsilon U_0}T_{zz},
\end{equation}
and
\begin{equation}
P=\frac{\varepsilon d}{U_0} p, \Sigma_n^{(ext)}=\frac{\varepsilon d}{U_0} T_n^{(ext)}, \Sigma_t^{(ext)}=\frac{d}{U}_0T_t^{(ext)}.
\end{equation}
Finally, the surface tension in the proper units is redefined as \cite{oron:1997} $\tilde{\sigma}=\varepsilon^3\sigma$.
Introducing this change of variables into the governing equations and boundary conditions, and keeping the dominant contribution to order ${\rm O}(\varepsilon^0)$, we find
\begin{eqnarray}
\partial_XU+\partial_ZW&=&0,\\
-\partial_XP+\partial_Z\Sigma_{XZ}&=&B_X,\label{Plubrication}\\
-\partial_ZP=-B_Z,\\
P|_{Z=H(X,T)}&=&-\tilde{\sigma}\partial^2_X H-\Sigma_n^{(ext)},\\
\Sigma_{XZ}|_{Z=H(X,T)}&=&\Sigma_t^{(ext)}+\partial_X\tilde{\sigma}.
\label{Plubrication2}
\end{eqnarray}
Integrating this system, one arrives to
\begin{eqnarray}
\partial_TH+\partial_X\int_0^{H(X,T)}U(X,Z,T)dZ&=&-J_{er},\label{kinematic}
\end{eqnarray}
where $J_{er}$ is the contribution from erosion in the scaled variables.

In order to solve (\ref{Plubrication})-(\ref{Plubrication2}), one needs some closure relation between the stress and the velocity. In our problem, we have split the stress into a body force and a (Newtonian) viscous tensor. For such a case, one arrives to a closed non-linear equation of the form
\begin{eqnarray}
\partial_TH&=&-\partial_X\Bigg[ \frac{H^3}{3}(-B_X+B_ZH_X+\tilde\sigma H_{XXX}+\partial_X\Sigma_n^{(ext)})\nonumber\\
&&-\frac{H^4}{8}\frac{d B_Z}{d\gamma }\frac{d\gamma}{d(H_X)} H_{XX}+\left(\Sigma^{(ext)}_t+\frac{d\tilde\sigma }{dX}\right)\frac{H^2}{2}\Bigg] -J_ {er} \nonumber
\label{lubricationeq}
\end{eqnarray}
with
\begin{eqnarray}
B_X&=&\frac{U_0}{d^2}f_E\Psi(\theta-\gamma)\sin\theta,\\
P(X,Z,T)&=&-B_Z(Z-H)+\tilde{\sigma}\partial^2_XH-T_n^{(ext)}.\\
B_Z&=&\frac{U_0}{d^2}f_E\Psi(\theta-\gamma)\cos\theta.
\end{eqnarray}

Hence, with $\Psi(\theta )=\cos\theta $ and $F_E=f_EU_0/d^2$,
\begin{eqnarray}
\partial_TH&=&-\partial_X\Bigg[ \frac{F_EH^3}{3}\left(\frac{-\cos\theta\sin\theta -\cos2\theta H_X+\sin\theta \cos\theta H_X^2}{(1+H_X^2)^{1/2}}\right)\nonumber \\
&&+\frac{H^3}{3}\tilde\sigma H_{XXX}-\frac{F_EH_{XX}H^4}{8}\frac{\sin\theta \cos\theta -\cos^2\theta H_X}{(1+H_X^2)^{3/2}}\Bigg]\\&&-J_ {er}.
\label{lubricationeq2}
\end{eqnarray}
Within our framework, an alternative equation can also be derived that features the same real part of the linear
dispersion relation, setting $B_{X,Z}=0$, $\Sigma_t^{(ext)}={\bf t}\cdot {\mathcal T} \cdot{\bf n}$, and $\Sigma_n^{(ext)}={\bf n}\cdot {\mathcal T} \cdot{\bf n}$, with $ \mathcal T $ as derived from Ref.\ \cite{spencer} for a two-dimensional coordinate system aligned with the ion beam,
\begin{eqnarray}
{\mathcal T} = F'_E
\left(
\begin{array}{cc}
-\cos \theta ^2+\nu \sin^2 \theta & -(1+\nu ) \cos \theta \sin \theta \\
-(1+\nu ) \cos \theta \sin \theta & \nu \cos^2 \theta -\sin^2 \theta
\end{array}
\right),
\end{eqnarray}
where $\nu=1$ for a two-dimensional incompressible fluid. Thus,
\begin{eqnarray}
\partial_TH&=&-\partial_X\Bigg[\frac{F'_EH^2}{2}\left(\frac{(H_X^2-1)\sin2\theta +2\cos 2\theta H_X}{1+H_X^2}\right)\nonumber\\
&&+\frac{H^3}{3}\left(\tilde\sigma H_{XXX}+F'_E\partial_X\left(\frac{(1-H_X^2)\cos 2\theta+2\sin 2\theta H_X}{1+H_X^2} \right)\right)\Bigg]-J_ {er}, 
\label{}
\end{eqnarray}
with $F_E\neq F'_E<0$ for compressive stress (the experimental case).

%}

Linearizing Eq.\ (\ref{lubricationeq}) and neglecting the effects of external stresses and the Marangoni term ($\partial_X\tilde{\sigma}$), we reproduce the shallow water dispersion relation given by Eq.\ (\ref{rewqshallow}).
The interest of this lubrication approach is that, provided an appropriate closure relation between the stress tensor and the velocity field is put into Eq.\ (\ref{Plubrication}), we obtain a non-linear equation for the surface height through Eq.\ (\ref{kinematic}), given Eq.\ (\ref{Plubrication}). Thus, nonlinear effects can be assessed that go beyond the analysis 
performed in Section \ref{la}.

\section{Conclusions}

In this work we have introduced a physical framework for surface pattern formation by IBS that is inspired in classic Fluid Mechanics. The strength of the theory is that it is based on general conservation laws, and that all the physics of the erosive process can be included into appropriate constitutive equations for the stress (body force) and boundary conditions. This approach allows to properly account for different mechanisms, such as viscous flow, erosion, or ion-induced strain (damage).

We have performed a linear stability analysis of the governing equations and shown that the forces related to the stress produced by the ion impacts indeed influence the morphological stability of the interface. For a specific choice of the angular dependence of the body force, $\Psi$, we have seen that some of the complexities of the morphological diagram for Si that have been unveiled recently, can be accounted for by our description. In this connection, purely erosive contributions indeed seem to play a minor role in the surface dynamics as compared with viscous flow.

Finally, the proposed hydrodynamic framework can be seen to generalize previous two-field formulations of IBS, to the case of a flowing layer of arbitrary thickness. Such phenomenological models have proved successful to obtain an effective equation describing the non-linear dynamics of the surface, reaching in some cases quantitative \cite{munoz-garcia:2010} and semiquantitative \cite{kim:2011b} agreement with experiments. On the other hand, our present framework does allow
for a systematic study of nonlinear effects through a lubrication theory approach.

The specific details of the constitutive equations that are appropriate for different experimental conditions will be the aim of future work. Also the general non-linear theory will be addressed through standard procedures~\cite{oron:1997} that should be able to account for nonlinear phenomena such as pattern coarsening, saturation of the roughness, or defect creation and/or annihilation.

\section*{Acknowledgments}

This work has been partially supported by MICINN (Spain) Grants No.\ FIS2009-12964-C05-01 and No.\ FIS2009-12964-C05-03. %Ministry of Science and Innovation.



%% References



\begin{thebibliography}{00}

\bibitem{cunningham:1960} R. L. Cunningham {\em et al.}, %P. Haymann, C. Lecomte, W. J. Moore, and J. J. Trillat,
J. Appl. Phys. {\bf 31}, 839 (1960).

\bibitem{navez:1962} M. Navez, C. Sella, and D. Chaperot, C. R. Acad. Sci. Paris {\bf 254}, 240 (1962).

\bibitem{gnaser:1998} H. Gnaser, {\it Low Energy Ion Irradiation of Solid Surfaces} (Springer, New York, 1998).

\bibitem{munoz-garcia:2010} J. Mu\~noz-Garc\'{\i}a, R. Gago, L V\'azquez, J. A. S\'anchez-Garc{\i}a, and Rodolfo Cuerno, Phys. Rev. Lett. {\bf 104}, 026101 (2010).

\bibitem{chan:2007_et_al} W. L. Chan and E. Chason, J. Appl. Phys. {\bf 101}, 121301 (2007).

\bibitem{munoz-garcia:2009} J. Mu\~noz-Garc\'{\i}a, L. V\'azquez, R. Cuerno, Jos\'e A. S\'anchez-Garc\'{\i}a, M. Castro, and R. Gago, in {\em Towards Functional Nanomaterials}, edited by Z. M. Wang (Springer, New York, 2009).

\bibitem{cuerno:2011} R. Cuerno, M. Castro, J. Mu\~noz-Garc\'{\i}a, R. Gago, and L. V\'azquez, Nucl. Instrum. Meth. Phys. Res. B {\bf 269}, 894 (2011).

%\bibitem{facsko:1999} S. Facsko {\em et al.}, %, T. Dekorsy, C. Koerdt, C. Trappe, H. Kurz, A. Vogt, and H. L. Hartnagel,
%Science {\bf 285}, 1551 (1999).
%
\bibitem{volkert:1993} C. A. Volkert, J. Appl. Phys. {\bf 74}, 7107 (1993).

\bibitem{mayr:2003} S. G. Mayr, Y. Ashkenazy, K. Albe, and R. S. Averback,
Phys. Rev. Lett. {\bf 90}, 055505 (2003).

\bibitem{bradley:1988} R. M. Bradley and J. M. E. Harper, J. Vac. Sci. Technol. A {\bf 6}, 2390 (1988).

\bibitem{cross:2009} M. Cross and H. Greenside, {\em Pattern Formation and Dynamics in Nonequilibrium Systems} (Cambridge University Press, Cambridge, 2009).

\bibitem{ozaydin:2005} G. Ozaydin, A. S. Ozcan, Y. Wang, K. F. Ludwig, H. Zhou, R. L. Headrick, and D. P. Siddons,
    Appl. Phys. Lett. {\bf 87} (2005) 163104.

\bibitem{madi:2008} C. S. Madi, B. Davidovitch, H. B. George, S. A. Norris, M. P. Brenner, and M. J. Aziz,
Phys. Rev. Lett. {\bf 101}, 246102 (2008).

\bibitem{madi:2009} C. S. Madi, H. B. George, and M. J. Aziz, J. Phys.: Condens. Matter {\bf 21}, 224010 (2009).

\bibitem{macko:2010} S. Macko, F. Frost, B. Ziberi, D. F. F\"orster, Th. Michely, Nanotechnology {\bf 21},
085301 (2010).

\bibitem{macko:2011} S. Macko, F. Frost, M. Engler, D. Hirsch, T. H\"oche, J. Grenzer, and T. Michely,
New J. Phys. {\bf 13}, 073017 (2011).

\bibitem{carter:1996} G. Carter and V. Vishnyakov, Phys. Rev. B {\bf 54}, 17647 (1996).

\bibitem{shenoy_et_al} V. Shenoy, W. L. Chan, and E. Chason, Phys. Rev. Lett. {\bf 98}, 256101 (2007).

\bibitem{leroy:2010} S. Le Roy, E. Sondergard, I. S. Nerbo, M. Kildemo, and M. Plapp, Phys. Rev. B {\bf 81}, 161401 (2010).

\bibitem{bradley:2010} R. M. Bradley and P. D. Shipman, Phys. Rev. Lett. {\bf 105}, 145501 (2010).

\bibitem{zhou:2010} J. Zhou and M. Lu, Phys. Rev. B {\bf 82}, 125404 (2010).

\bibitem{makeev:2002} M. Makeev, R. Cuerno, and A.-L. Barab\'asi, Nucl. Inst. Meth. Phys. Res. B {\bf 197}, 185 (2002).

\bibitem{facsko:2004} S. Facsko, T. Bobek, A. Stahl, H. Kurz and T. Dekorsy, Phys. Rev. B {\bf 69} 153412 (2004).

\bibitem{davidovitch:2007} B. Davidovich, M. Aziz, and M. Brenner, Phys. Rev. B {\bf 76}, 205420 (2007).

\bibitem{castro:2005} M. Castro,  R. Cuerno, L. V\'azquez, and R. Gago, Phys. Rev. Lett. {\bf 94}, 016102 (2005).

\bibitem{munoz:2006} J.\ Mu\~noz-Garc\'{\i}a, M.\ Castro, R.\ Cuerno, Phys. Rev. Lett. {\bf 96}, 086101 (2006).

\bibitem{munoz:2008} J.\ Mu\~noz-Garc\'{\i}a, R.\ Cuerno, and M.\ Castro, Phys. Rev. B {\bf 78}, 205408 (2008).

\bibitem{munoz:2009b} J.\ Mu\~noz-Garc\'{\i}a, R.\ Cuerno, and M.\ Castro, J. Phys. Condens. Matter {\bf 21}, 224020 (2009).

\bibitem{pelaz:2004} L. Pelaz, L. A. Marqu\'es, and J. Barbolla, J. Appl. Phys. {\bf 96}, 5947 (2004).

\bibitem{kalyanasaundaram:2008} N. Kalyanasundaram, M. Wood, J. B. Freund, and H. T. Johnson,
Mech. Res. Comm. {\bf 35}, 50 (2008).

\bibitem{moseler:2005} M. Moseler, P. Gumbsch, C. Casiraghi, A. C. Ferrari, and J. Robertson,
Science {\bf 309}, 1545 (2005).

\bibitem{moore:2004} M. C. Moore, N. Kalyanasundaram, J. B. Freund, and H. T. Johnson,
Nucl. Inst. Meth. Phys. Res. B {\bf 225}, 241 (2004).

\bibitem{oron:1997} A. Oron, S. H. Davis, and S. G. Bankoff, Rev. Mod. Phys. {\bf 69}, 931 (1997).

\bibitem{Kree} R.\ Kree, oral communication at International Conference on Ion-Beam Induced Nanopatterning of Materials (IINM-2011), 06-10 February, 2011.

\bibitem{arxiv} M. Castro and R. Cuerno, arXiv:1007.2144v1 (2011).

\bibitem{orchard:1962} S. E. Orchard, Appl. Sci. Res. {\bf 11A}, 451 (1962).

\bibitem{mullins:1959} W. W. Mullins, J. Appl. Phys. {\bf 30}, 77 (1959).

\bibitem{chason:1994} E. Chason, T. M. Mayer, B. K. Kellerman, D. T. Mcilroy, and A. J. Howard,
Phys. Rev. Lett. {\bf 72}, 3040 (1994).

\bibitem{frost:2009} F. Frost, R. Fechner, B. Ziberi, J. V\"ollner, D. Flamm, and A. Schindler,
J. Phys. Condens. Matter {\bf 21}, 224026 (2009).

\bibitem{norris:2011} S. A. Norris, J. Samela, L. Bukonte, M. Backman, F. Djurabekova, K. Nordlund, C. S. Madi,
M. P. Brenner, and M. J. Aziz, Nat. Comm. {\bf 2}, 276 (2011).

\bibitem{kim:2011} S.-P Kim, B.-H. Kim, H. Kim, K.-R. Lee, Y.-C. Chung, J. Seo, and J.-S. Kim, Nucl. Instrum. Meth. Phys. Res. B {\bf 269}, 2605 (2011).

\bibitem{vauth:2007} S. Vauth and S. G. Mayr, Phys. Rev. B {\bf 75}, 224107 (2007).

\bibitem{aste} T. Aste and U. Valbusa, Physica A {\bf 332}, 548
(2004); New J. Phys. 7, 122 (2005).

\bibitem{spencer} B.J. Spencer, S.H. Davis and P.W. Voorhees, Phys. Rev. B 47 (1993) 9760.


\bibitem{kim:2011b} J-H. Kim, N.-B. Ha, J.-S. Kim, M. Joe, K.-R. Lee, and R. Cuerno,
Nanotechnology {\bf 22}, 285301 (2011).

\end{thebibliography}



\appendix
\section{Solutions of the linearized governing equations}
\label{app:a}

Expressions for $P_1(z)$, $u_1(z)$ and $w_1(z)$ in Eqs.\ (\ref{Pexpansion}) to (\ref{wexpansion}) of the main text are
\begin{eqnarray}
&&P_1(z)=\frac{1}{2 \bigg(1+2 d^2 q^2+\cosh(2 d q)\bigg)}e^{-q (2 d+z)} \Bigg[\nonumber \\
&&q^2 \bigg(1+e^{2 d q} \bigg(1+2 d q+e^{2 q z} \bigg(1+e^{2 d q}-2 d q\bigg)\bigg)\bigg) \sigma +2 e^{q (2 d+z)} (f_E \cos(\theta ) (\cosh(q z)+\nonumber \\
&&\cosh(q (2 d+z))-2 d q \sinh(q z))+i f_E \sin(\theta ) (2 d q \cosh(q z)+\sinh(q z)+\nonumber\\
&& \sinh(q (2 d+z)))) \Psi (\theta )+\bigg(i f_E \bigg(1-e^{d q} \bigg(2+2 e^{2 d q}-2 e^{2 q z}+e^{3 d q+2 q z}-\nonumber\\
&&4 d q+e^{d q} (-1+2 d q)-2 e^{2 q (d+z)} (1+2 d q)e^{q (d+2 z)} (1+2 d q)\bigg)\bigg) \cos(\theta )-\nonumber \\
&&f_E \bigg(1-e^{q z}+e^{2 q (2 d+z)}-e^{q (4 d+z)}+e^{2 q (d+z)} (1-2 d q)+e^{2 d q} (1+2 d q)-\nonumber\\
&&2 e^{q (2 d+z)} \bigg(1+2 d^2 q^2\bigg)\bigg) \sin(\theta )\bigg) \Psi '(\theta )\Bigg],
\label{P1}
\end{eqnarray}

\begin{eqnarray}
&&u_1(z)=\frac{1}{2 q  \mu\bigg(1+e^{4 d q}+e^{2 d q} \bigg(2+4 d^2 q^2\bigg)\bigg) }e^{-q z} \Bigg[\nonumber\\
&&
i q^3 \bigg(z-e^{2 q (2 d+z)} z+e^{2 q (d+z)} (-z+2 d q (d+z))+e^{2 d q} (z+2 d q (d+z))\bigg) \sigma +\nonumber\\
&&f_E \bigg(i q \bigg(z-e^{2 q (2 d+z)} z+e^{2 q (d+z)} (-z+2 d q (d+z))+e^{2 d q} (z+2 d q (d+z))\bigg) \cos(\theta )+\nonumber\\
&& \bigg(-1+q z+e^{2 q (2 d+z)} (1+q z)+e^{2 d q} (-1+q (z-2 d (-1+q (d+z))))+\nonumber\\
&&e^{2 q (d+z)} (1+q (z+2 d (1+q (d+z))))\bigg) \sin(\theta )\bigg) \Psi (\theta )+\nonumber\\
&&f_E \bigg(\bigg(1-q z+2 e^{3 d q} q (d+z)+2 e^{q (d+2 z)} q (d+z)-e^{2 q (2 d+z)} (1+q z)\nonumber\\
&&-2 e^{d q} q (-d-z+2 d q z)+2 e^{3 d q+2 q z} q (d+z+2 d q z)+e^{2 d q} (1+2 d q (-1+d q)\nonumber\\
&&+q (-1+2 d q) z)-e^{2 q (d+z)} (1+q (z+2 d (1+q (d+z))))\bigg) \cos(\theta )\nonumber\\
&&-i q \big(z-e^{2 q (2 d+z)} z+e^{2 q (d+z)} (-z+2 d q (d+z))+e^{2 d q} (z+\nonumber \\
&&2 d q (d+z))\big) \sin(\theta )\bigg) \Psi '(\theta ) \Bigg],
\label{u1}
\end{eqnarray}
and
\begin{eqnarray}
&&w_1(z)=\frac{1}{2 q \mu \bigg(1+e^{4 d q}+e^{2 d q} \bigg(2+4 d^2 q^2\bigg)\bigg) }e^{-q z} \Bigg[\nonumber \\
&&q^2 \bigg(1+q z+e^{2 q (2 d+z)} (-1+q z)+e^{2 q (d+z)} (-1+q (z-2 d (-1+q (d+z))))+\nonumber \\
&&e^{2 d q} (1+q (z+2 d (1+q (d+z))))\bigg) \sigma +\bigg(f_E \bigg(1+q z+e^{2 q (2 d+z)} (-1+q z)+\nonumber \\
&&e^{2 q (d+z)} (-1+q (z-2 d (-1+q (d+z))))+e^{2 d q} (1+q (z+2 d (1+q (d+z))))\bigg) \cos(\theta)\nonumber \\
&&+i f_E q \bigg(-z+e^{2 q (2 d+z)} z+e^{2 d q} (-z+2 d q (d+z))+e^{2 q (d+z)} (z+\nonumber \\
&&2 d q (d+z))\bigg) \sin(\theta)\bigg) \Psi (\theta)+\bigg(i f_E \bigg(2 e^{q z}+2 e^{q (4 d+z)}+e^{q (2 d+z)} \bigg(4+8 d^2 q^2\bigg)+\nonumber \\
&&q z-e^{2 q (2 d+z)} q z+ 2 e^{d q} (-1+d q+q (-1+2 d q) z)+2 e^{q (d+2 z)} (-1+q (d+z))-\nonumber \\
&&2 e^{3 d q} (1+q (d+z))-e^{2 d q} q (-z+2 d q (d+z))-e^{2 q (d+z)} q (z+2 d q (d+z))+\nonumber \\
&& 2 e^{3 d q+2 q z} (-1+q z+d q (-1+2 q z))\bigg) \cos(\theta)+f_E \sin(\theta) \bigg(-1-q z+e^{2 q (2 d+z)} (1-q z)+\nonumber \\
&& 2 e^{q (2 d+z)} \bigg(-q (2 d+z) \cosh(q z)+\bigg(1+2 d q^2 (d+z)\bigg) \sinh(q z)\bigg)\bigg)\bigg) \Psi '(\theta)\Bigg].
\label{w1}
\end{eqnarray}

\section*{References´}

\end{document}